\documentclass{iopart}

 \usepackage{graphicx}
 \usepackage{bm}

\begin{document}

\title{Medium-modified Casimir forces}
 \author{M. S. Toma\v s}
 \address{Rudjer Bo\v skovi\' c Institute, P. O. B. 180,
 10002 Zagreb, Croatia}
 \ead{tomas@thphys.irb.hr}

 \begin{abstract}
We argue that the results for the vacuum forces on a slab and on
an atom embedded in a magnetodielectric medium near a mirror,
obtained using a recently suggested Lorentz-force approach to the
Casimir effect, are equivalent to the corresponding results
obtained in a traditional way. We also derive a general expression
for the atom-atom force in a medium and extend a few classical
results concerning this force in vacuum and dielectrics to
magnetodielectric systems. This, for example, reveals that the
(repulsive) interaction between atoms of different polarizability
type is at small distances unaffected by a (weakly polarizable)
medium.
\end{abstract}


Although modifications of the Casimir and van der Waals forces due
to the presence of a medium between the interacting objects were
the subject of interest for a long time
\cite{Dzy,Schw,Zhou,Abr,Mil}, this issue is still of great
importance owing to the dominant role of these forces at small
distances and to the rapid progress in micro and nanotechnologies.
A common way of extending the Lifshitz theory \cite{Lif} of the
Casimir effect \cite{Cas} to material cavities is to (eventually)
employ the Minkovski stress tensor
$T_{ij}=\frac{1}{4\pi}\left<D_iE_j+H_iB_j-\frac{1}{2}
 \left({\bf D}\cdot{\bf E}+{\bf H}\cdot{\bf B}\right)\delta_{ij}\right>$
when calculating the force \cite{Dzy,Schw,Abr,Tom02}. Recently,
however, a Lorentz-force approach to the Casimir effect was
suggested \cite{Raa04} (see also \cite{Obu}) in which the relevant
stress tensor is of the form (brackets denote the average with
respect to fluctuations and we use the standard notation for the
macroscopic field operators)
\begin{equation}
 T^{(L)}_{ij}=T_{ij}-\left<P_iE_j-M_iB_j-\frac{1}{2}\left({\bf P}\cdot{\bf E}-
 {\bf M}\cdot{\bf B}\right)\delta_{ij}\right>.
 \label{T}
 \end{equation}
In this work, we reinterpret the results for the vacuum forces on
a slab and on an atom embedded in a semi-infinite
magnetodielectric cavity (see Figure 1) as recently obtained using
$T^{(L)}_{ij}$ \cite{Tom05,Tom052} and argue their equivalence to
the corresponding results obtained using $T_{ij}$. Also, we derive
a general expression for the atom-atom force in a medium and
extend a few well-known results for this force to
magnetodielectric systems.

\begin{figure}[htb]

 \begin{center}
 \resizebox{8cm}{!}{\includegraphics{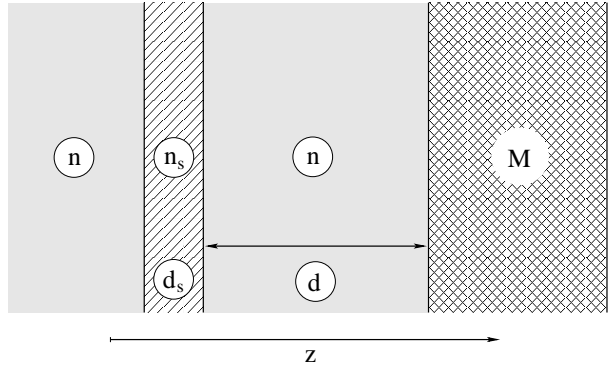}}
 \end{center}
 \caption{A slab in front of a mirror shown schematically. The
 (complex) refraction index of the slab is
 $n_s(\omega)=\sqrt{\varepsilon_s(\omega)\mu_s(\omega)}$ and that
 of the surrounding medium is $n(\omega)=\sqrt{\varepsilon(\omega)\mu(\omega)}$.
 The mirror is described by its reflection coefficients
 $R^q(\omega,k)$, with $k$ being the in-plane wave vector of a wave.
 The arrow indicates the direction of the force on the slab.\label{sys}}
\end{figure}

When calculating $T^{(L)}_{ij}$ for planar geometry \cite{Raa04},
the zero-temperature force on the slab per unit area in the
configuration of Figure \ref{sys} can be written as \cite{Tom05}

\numparts
 \begin{equation}
f^{(L)}(d)=f(d)+\tilde{f}(d),
 \end{equation}
\begin{equation}
\label{f}
 f(d)=\frac{\hbar}{2\pi^2}\int_0^\infty
d\xi \int^\infty_0dkk\kappa \sum_{q=p,s}\left[\frac{1}
{(r^qR^q\rme^{-2\kappa d})^{-1}-1}\right](i\xi,k),
\end{equation}
\begin{eqnarray}
\label{tf}
\fl \tilde{f}(d)&=\frac{\hbar}{2\pi^2}\int_0^\infty
d\xi \int^\infty_0dkk\kappa
\sum_{q=p,s}\left[\frac{(\varepsilon^{-1}-1)\delta_{qp}+(\mu-1)\delta_{qs}}
{(r^qR^q\rme^{-2\kappa d})^{-1}-1}\right](i\xi,k)\nonumber\\
\fl &+\frac{\hbar}{8\pi^2c^2}\int_0^\infty
 d\xi\xi^2\mu(n^2-1)\int^\infty_0\frac{dkk}{\kappa}
\sum_{q=p,s}\Delta_q
\left[\frac{(1+r^q)^2-{t^q}^2}{(R^q\rme^{-2\kappa d})^{-1}
-r^q}\right](i\xi,k),
 \end{eqnarray}
 \endnumparts
corresponding to the decomposition of the stress tensor in
(\ref{T}). Here $\kappa(\xi,k)=\sqrt{n^2(i\xi)\xi^2/c^2+k^2}$ is
the perpendicular wave vector in the cavity at the imaginary
frequency, $\Delta_q=\delta_{qp}-\delta_{qs}$, $r^q$ and $t^q$ are
Fresnel coefficients for the (symmetrically bounded) slab and
$R^q$ are those for the mirror.
As seen, the first term in $\tilde{f}$ may be combined with the
traditional (Minkowski) force $f$ to form a medium-screened
Casimir force, with the contributions of TM and TE polarized waves
scaled, respectively, by $1/\varepsilon$ and $\mu$, whereas the
remaining term in $\tilde{f}$ may be regarded as a medium-assisted
force \cite{Tom05}. However, owing to this (additional) screening
of the force, it turns out that the medium-screened van der Waals
and Casimir forces have a rather unusual dependence on the
material parameters of the system \cite{Tom052} and, therefore,
this formal combination demands reconsideration. Here, we
interpret (2) by recalling that, according to the $T\neq 0$
quantum-field-theoretical approach to the Casimir force, the
Minkowski stress tensor $T_{ij}$ corresponds to the effective
stress tensor in a medium which is in mechanical equilibrium
\cite{Dzy,Abr,Pit}. If so, the (unbalanced) second term in
(\ref{T}), which leads to $\tilde{f}$, gives a force on the
medium. Therefore, the force on the slab is given solely by the
traditional force $f$ in (2), whereas $\tilde{f}$ describes a
force on the medium. Indeed, note that $\tilde{f}$ vanishes when
there is no medium, $\varepsilon=\mu=1$, and is nonzero when there
is no slab $n_s=n$ [$r^q=0$ and $t^q=e^{-\kappa d_s}$ in
(\ref{tf})], in this case clearly representing the force on a
layer of the medium.

The force on an atom near a mirror can be obtained from (2) by
assuming that the slab consists of a thin ($d_s\ll d$) layer of
the cavity medium with a small number of atoms $A$ embedded in it
\cite{Tom052}. Then, from the Lorentz-Lorenz (Clausius-Mossotti)
relation \cite{AsLa}, it follows that
\begin{equation}
 \varepsilon_s=\varepsilon+4\pi
N_A\alpha^A_e,\;\;\;\alpha^A_e=\alpha^A_{e0}(\frac{\varepsilon+2}{3})^2
\left[1+{\cal O}(N_A\alpha^A_{e0})\right],
\end{equation}
where $N_A$ is the number density and $\alpha^A_{e0}$ the electric
vacuum polarizability of embedded atoms. Assuming a similar
relation between $\mu_s$, $\mu$ and the effective atomic magnetic
polarizability $\alpha^A_m$, for the force on an {\it embedded}
atom at distance $d$ from the mirror we find (for details, see the
derivation of Equation (14) in \cite{Tom052} \footnote{Owing to a
lapse, the replacements $\alpha_e\rightarrow\alpha_e/\varepsilon$
and $\alpha_m\rightarrow\alpha_m/\mu$ must be made in the last
term of this equation.})

\numparts\label{faL}
\begin{equation}\label{fL}
f^{(L)}_A(d)=f_A(d)+\tilde{f}_A(d),
\end{equation}
\begin{eqnarray}
 \label{fa}
\fl f_A(d)&=\frac{\hbar}{\pi c^2}\int_0^\infty d\xi\xi^2
\int^\infty_0dkk \rme^{-2\kappa d} \left\{\left[\alpha^A_e
\left(2\frac{\kappa^2c^2}{\varepsilon\xi^2}-\mu\right)
-\alpha^A_m\varepsilon\right]R^p\right.
\nonumber\\
\fl &+\left.\left[\alpha^A_m
\left(2\frac{\kappa^2c^2}{\mu\xi^2}-\varepsilon\right)
-\alpha^A_e\mu\right]R^s\right\}(i\xi,k),
\end{eqnarray}
\begin{eqnarray}
\label{tfa} \fl \tilde{f}_A(d)&=\frac{\hbar}{\pi c^2}\int_0^\infty
d\xi\xi^2 \int^\infty_0dkk e^{-2\kappa d}
\left\{\left(\frac{1}{\varepsilon}-1\right)\left[\alpha_e
\left(2\frac{\kappa^2c^2}{\varepsilon\xi^2}-\mu\right)
-\alpha_m\varepsilon\right]R^p\right.\\
\fl &\left.+(\mu-1)\left[\alpha_m
\left(2\frac{\kappa^2c^2}{\mu\xi^2}-\varepsilon\right)
-\alpha_e\mu\right]R^s+\left(\mu-\frac{1}{\varepsilon}\right)
\left[\alpha_e\mu R^p-\alpha_m\varepsilon
R^s\right]\right\}(i\xi,k).\nonumber
\end{eqnarray}
\endnumparts
As before, this can be combined to give a medium-screened and a
medium-assisted force on the atom \cite{Tom052}. However,
according to the interpretation adopted here, the true force on
the atom is $f_A$, whereas $\tilde{f}_A$ may be regarded as an
atom-induced force on the medium (per atom). We note that
(\ref{fa}) extends (in different directions) previous results for
the atom-mirror force in various circumstances
\cite{Schw,Zhou,CP,Boy,Buh} by fully accounting for the magnetic
properties of the system.

\Eref{fa} enables one to calculate the force between two atoms in
a magnetodielectric medium. This force can be found by assuming a
single-medium mirror consisting of the cavity medium with a small
number of, say, type $B$ atoms embedded in it, so that
$\varepsilon_m=\varepsilon+4\pi N_B\alpha^B_e$ and $\mu_m=\mu+4\pi
N_B\alpha^B_m$. In this case, the reflection coefficients of the
mirror can be approximated by \cite{Tom052}
\begin{equation}
\label{Rp}
R^p(i\xi,k)\simeq \frac{\pi N_B\xi^2}{\kappa^2 c^2}
\left[\alpha^B_e\mu(\frac{2\kappa^2c^2}{n^2
\xi^2}-1)-\alpha^B_m\varepsilon\right]
\end{equation}
and $R^s=R^p(\varepsilon\leftrightarrow\mu,\;
\alpha^B_e\leftrightarrow\alpha^B_m)$. Also, the potential energy
of the atom $U_A(d)=-\int^d_\infty \rmd l f_A(l)$ is given by
$U_A(d)=N_B\int_{z\geq d}\rmd^3{\bf r}U_{AB}(r)$, where ${\bf
r}={\bf r}_B-{\bf r}_A$ and $U_{AB}(r)$ is the interaction energy
between the atoms $A$ and $B$. Using the identity $\exp(-2\kappa
d)=(2\kappa^2/\pi)\int_{z\geq d}\rmd^3{\bf r}\exp(-2\kappa r)/r$
and combining these relations, from (\ref{fa}) and (\ref{Rp}) we
obtain the extension of the Feinberg and Sucher formula \cite{Fei}
to material systems
\begin{eqnarray}\label{UAB}
\fl U_{AB}(r)=&\frac{\hbar}{16\pi r^6}\int_0^\infty \rmd\xi
\rme^{-2n\xi r/c}F(\frac{2n\xi r}{c})
\left(\frac{\alpha^A_e\alpha^B_e}{\varepsilon^2}+
\frac{\alpha^A_m\alpha^B_m}{\mu^2}\right)\nonumber\\
\fl &-\frac{\hbar}{4\pi c^2 r^4}\int_0^\infty \rmd\xi \xi^2
\rme^{-2n\xi r/c}G(\frac{2n\xi
r}{c})(\alpha^A_e\alpha^B_m+\alpha^A_m\alpha^B_e),
\end{eqnarray}
where $F(x)=x^4+4x^3+20x^2+48x+48$ and $G(x)=(x+2)^2$.

At small distances between the atoms, $r\ll c/\Omega$ ($\Omega$ is
an effective upper limit in the integration over $\xi$
\cite{Buh}), we may let $\exp(-2n\xi r/c)\simeq 1$, $F(2n\xi
r/c)\simeq 48$ and $G(2n\xi r/c)\simeq 4$ in (\ref{UAB}). This
gives for the atom-atom force $f_{AB}=-\rmd U_{AB}/\rmd r$ at
small distances
\begin{equation}
\fl f_{AB}(r)=\frac{18\hbar}{\pi r^7}\int_0^\infty d\xi
\left(\frac{\alpha^A_e\alpha^B_e}{\varepsilon^2}+
\frac{\alpha^A_m\alpha^B_m}{\mu^2}\right)-\frac{4\hbar}{\pi
c^2r^5}\int_0^\infty d\xi\xi^2
(\alpha^A_e\alpha^B_m+\alpha^A_m\alpha^B_e), \label{faas}
\end{equation}
which generalizes the well-known results for the van der
Waals-London force \cite{Dzy,Lon,Fei}. We observe that, whereas
the force between the atoms of the same polarizability type is
strongly affected by the surrounding medium, the (repulsive) force
between the atoms of different polarizability type (in
weakly-polarizable media) remains the same as in vacuum. At large
distances between the atoms, the main contribution to the integral
in (\ref{UAB}) comes from the $\xi\simeq 0$ region. Approximating
the $\xi$-dependent quantities by their static values (denoted by
the subscript $0$) and then performing the integration, we arrive
at
\begin{equation}
\fl f_{AB}(r)=\frac{7\hbar c}{4\pi n^5_0 r^8}
\left[23\left(\alpha^A_e\alpha^B_e\mu^2+
\alpha^A_m\alpha^B_m\varepsilon^2\right)
-7\left(\alpha^A_e\alpha^B_m+
\alpha^A_m\alpha^B_e\right)\varepsilon\mu\right]_0, \label{faal}
\end{equation}
which gives the medium corrections to the retarded atom-atom force
\cite{CP,Fei} and extends the previous considerations of these
corrections \cite{Dzy,Abr} to magnetodielectric systems. As seen,
the repulsive component of the force is in a (weakly-polarizable)
medium simply scaled by $n^{-3}_0$, whereas the modification of
its attractive part is more complex.

We end this short discussion by noting that the symmetry of
(\ref{fa}) under the transformation:
$\alpha_e\leftrightarrow\alpha_m$, $\varepsilon\leftrightarrow\mu$
and $R^p\leftrightarrow R^s$ [and consequently that of
(\ref{UAB})-(\ref{faal}) with respect to the replacements
$\alpha_e\leftrightarrow\alpha_m$ and
$\varepsilon\leftrightarrow\mu$] is a consequence of the
invariance of the Minkowski stress tensor $T_{ij}$ with respect to
a duality transformation, e.g. ${\bf D}\rightarrow {\bf B}$, ${\bf
E}\rightarrow {\bf H}$, ${\bf B}\rightarrow -{\bf D}$ and ${\bf
H}\rightarrow -{\bf E}$ \cite{Boy}. This symmetry is lost in the
second term of $T^{(L)}_{ij}$, so that the force $\tilde{f}_A$ is
not invariant under the replacement of the electric and magnetic
quantities. Since the medium-screened atom-mirror and atom-atom
forces are combinations of the corresponding $f$'s and
$\tilde{f}$'s, this explains the unusual (asymmetric) medium
effects on these forces found in \cite{Tom052}.

In conclusion, when properly interpreted, the result for the
vacuum force on an object (a slab or an atom) embedded in a medium
near a mirror obtained using the Lorentz-force approach to the
Casimir effect agrees with that obtained in a traditional way.
Extensions of the well-known results for the atom-atom force to
magnetodielectric systems reveals that the (repulsive) interaction
between atoms of different polarizability type remains (in
weakly-polarizable media) the same as in vacuum at small and is
scaled by $n^{-3}_0$ at large distances. Medium corrections to the
force between atoms of the same polarizability type are more
complex and depend on the polarizability type of atoms.\\

\noindent The author thanks to C. Raabe and anonymous referees for
constructive criticism and suggestions. This work was supported in
part by the Ministry of Science and Technology of the Republic of
Croatia under contract No. 0098001.

\end{document}